\documentclass[nofootinbib,aps,prd,twocolumn,showpacs,preprintnumbers]{revtex4}
\usepackage{graphics}
\def \beq{\begin{equation}}
\def \eeq{\end{equation}}
\def \beqa{\begin{eqnarray}}
\def \eeqa{\end{eqnarray}}
\def\etc{{\sl etc.\/}}
\def\ie{{\sl i.\ e.\/}}
\def \a{c}

\def\etal{{\sl et al.\/}}
\def\jhep{{\sl J.\ H.\ E.\ P.\/}}
\begin{document}
\title{Screening correlators with chiral Fermions}
\author{R.\ V.\ \surname{Gavai}}
\email{gavai@tifr.res.in}
\affiliation{Department of Theoretical Physics, Tata Institute of Fundamental
         Research,\\ Homi Bhabha Road, Mumbai 400005, India.}
\author{Sourendu \surname{Gupta}}
\email{sgupta@tifr.res.in}
\affiliation{Department of Theoretical Physics, Tata Institute of Fundamental
         Research,\\ Homi Bhabha Road, Mumbai 400005, India.}
\author{R.\ \surname{Lacaze}}
\email{Robert.Lacaze@cea.fr}
\affiliation{Service de Physique Theorique, CEA Saclay,\\
         F-91191 Gif-sur-Yvette Cedex, France.}

\begin{abstract}
We study screening correlators of quark-antiquark composites at
$T=2T_c$, where $T_c$ is the QCD phase transition temperature, using
overlap quarks in the quenched approximation of lattice QCD. As
the lattice spacing is changed from $1/4T$ to $a=1/6T$ and $1/8T$,
we find that screening correlators change little, in contrast with
the situation for other types of lattice fermions.  All correlators
are close to the ideal gas prediction at small separations. The
long distance falloff is clearly exponential, showing that a
parametrization by a single screening length is possible at distances
$z\ge1/T$. The correlator corresponding to the thermal vector
is close to the ideal gas value at all distances, whereas that
for the thermal scalar deviates at large distances. This is
examined through the screening lengths and momentum space correlators.
There is strong evidence that the screening transfer matrix does
not have reflection positivity.
\end{abstract}
\preprint{TIFR/TH/06-xx, t06/rxx, hep-lat/yymmnnn}

\maketitle

\section{Introduction}

Current experiments at the Brookhaven RHIC and another to be started
soon at the CERN LHC are engaged in creating and understanding the
phase of matter called the quark gluon plasma. The existence of this
phase was predicted by lattice QCD, and this technique has been used
to study many properties of the plasma. In spite of this, much remains
to be done. One of the major questions, that which concerns us here,
is the nature of the interactions between quasiparticle excitations.

Decisive information on flavoured quasiparticles has come from a recent
study of the linkage between quantum numbers \cite{linkage}.  A rapid
and clear change in the linkage between quantum numbers was seen across
the QCD phase transition. In the high temperature phase flavour quantum
numbers such as baryon number, charge, strangeness \etc, are linked to
each other in ways that suggest that the light flavoured excitations at
finite temperature QCD include quarks.

At the same time it is clear that the interactions between these
quasiparticles cannot be small. At length scales of order $1/T$,
the QCD coupling, $g$, is of order 1 \cite{precise}, leading to well
known problems. For example, the Debye screening length of gluons is
more complicated than one expects in perturbation theory \cite{nadkarni}
and contains pieces which are entirely non-perturbative \cite{yaffe}.
The non-perturbative pieces have been isolated and studied on the lattice \cite{hels}.
Interestingly, when one studies the screening of glueball-like quantities,
\ie, spatial correlations of colour singlet operators made out of gauge
fields, then a surprising simplicity arises above about $1.25T_c$. Taking
two quantum number channels which can be constructed by a minimum of two
and three gluon operators, the ratio of the screening masses is seen to
be close to 3/2 \cite{saumen}.

There are parallels in the study of hadron-like screening correlators in
the QCD plasma. Parity doubling was seen in the high-temperature phase
of QCD in the first study of these correlators using staggered quarks
\cite{detar}. Further,
this first study already showed that the screening mass from the
baryon-like correlator was 3/2 times the screening mass from some of the
meson-like correlators.  A finite size scaling study using staggered
quarks \cite{mtc} showed that a particular combination of the vector
and axial-vector meson-like correlators was very close to the free-field
theory (\ie, ideal gas) predictions.  This behaviour is generic, being
seen in quenched \cite{quenched} and dynamical QCD with two \cite{two}
and four \cite{four} flavours of staggered quarks as well as with Wilson
quarks \cite{wilson}.  Extrapolation to the continuum using staggered
quarks \cite{conti} showed that the remaining screening masses approached
their ideal-gas values. However, the correlators differed strongly from
the ideal-gas correlators at small distances.  This was also seen in a
later study of screening correlators with Wilson quarks \cite{edwin}.

Overlap quarks \cite{neu} have the advantage of preserving all chiral
symmetries on the lattice for any number of massless flavors of quarks
\cite{luescher}.  This is in contrast to other formulations, such
as Wilson's, which break all chiral symmetries, or the staggered,
which break them partially.  Since the number of pions and
their nature is intimately related to the actually realized chiral
symmetry on the lattice, one expects any realization of chiral
quarks on the lattice to provide insight into this question. 

In our earlier study of screening correlators using overlap quarks
\cite{earlier} with lattice spacing $a=1/4T$, the correlation functions
were found to decay exponentially at large separation, and showed none of
the fine structure that plague staggered and Wilson quarks. As a result,
the screening mass was a good parametrization of the screening correlator
at large distances.  The screening masses in all channels were closer
to the expected weak-coupling continuum limit for overlap quarks than
staggered and Wilson. However, that from the (would be zero-temperature)
scalar and pseudoscalar was lower by about 10\% than the others.

In this paper we extract complete information on meson-like screening
correlators by extending the analysis of \cite{saumen,symm} to
overlap quarks. At the same time, we extend the analysis towards the
continuum limit by using lattice spacings of $a=1/6T$ and $1/8T$.
With these two inputs we are able to resolve all the currently
outstanding questions on screening correlators. Parts of these results
were presented in \cite{pos}.

\section{Symmetries}

\begin{table}[t]
\begin{center}\begin{tabular}{|c|c|c|c|c|}
\hline
  &\multicolumn{2}{c|}{$T=0$}&\multicolumn{2}{c|}{$T>0$}\\
\hline
  & continuum & lattice & continuum & lattice \\
\hline
  & $O(3)\times Z_2(P)$ & $\quad O_h\quad $ 
  & $O(2)\times Z_2({\cal T})$ & $\quad D_4^h\quad $\\
\hline
S  & $0^+$ & $A_1^+$ & $0^+$ & $A_1^+$ \\
PS & $0^-$ & $A_1^-$ & $0^-$ & $A_1^-$ \\
V  & $1^+$ & $F_1^+$ & $0^-$ & $A_1^-$ \\
   &       &         & $1^+$   & $E^+$   \\
AV & $1^-$ & $F_1^-$ & $0^+$ & $A_1^+$ \\
   &       &         & $1^-$   & $E^-$   \\
\hline
\end{tabular}\end{center}
\caption{The break up of irreps under the successive breakings of the
   symmetries of the transfer matrix. The $A_1^\pm$ components of the
   V/AV correspond to the $t$ polarization and the $E^\pm$ to the
   $x$ and $y$ polarizations. All states also carry a label for charge
   conjugation, $C$. This has been dropped in this paper since we consider
   only the $C=1$ states.}
\label{tb.symm}\end{table}

We study correlation functions of colour singlet operators constructed
with a quark and an antiquark---
\beqa
\nonumber
   C_\Gamma(z) &=& \langle M_\Gamma(0) M_\Gamma^\dag(z)\rangle,
     \qquad{\rm where}\\
   M_\Gamma(z) &=& \frac1{N_tN_s^2}
    \sum_{xyt}\overline\psi(txyz)\Gamma\psi(txyz),
\label{corr}\eeqa
$\psi$ is a Dirac spinor, $\Gamma$ is a Dirac matrix, the angular
brackets denote averaging over gauge configurations and the quark
bilinear $M_\Gamma$ is projected to zero momentum in the slice
orthogonal to $z$ by the summation.  By choosing different Dirac
matrices, $\Gamma$, one explores different quantum numbers, and the
usual nomenclature is explained in Table \ref{tb.symm}.
Note that the correlators projected on zero momentum ($k_x=k_y=k_t=0$)
have mass dimension 3, and hence the quantity $a^3 C(z)$ is dimensionless.

Since we work at finite temperature, the Dirac operator is defined
with antiperiodic boundary conditions on $t$ ($1\le t\le N_t$). We
chose to impose periodic boundary conditions in the spatial directions
($1\le x,y\le N_s$ and $1\le z\le N_z$ with $N_s\le N_z$). Later
we shall have occasion to use the aspect ratio $\zeta=N_z/N_s$.
The sum over $x$, $y$ and $t$ in eq.\ (\ref{corr}) accomplishes a
projection on to zero total momentum in these three directions.

Due to the inequivalence of the spatial directions $x$ and $y$ with
the Euclidean time direction $t$ at finite temperature, a slice of
the lattice orthogonal to the $z$ direction differ at zero and
finite temperature. These are part of the symmetries of the transfer
matrix.\footnote{The remaining symmetries are flavour symmetries.
For staggered quarks the analysis is complicated \cite{symm} by the
mixing of flavour and spacetime symmetries.} Therefore the
classification of operators $M_\Gamma$ differ at zero and finite
temperature.  This is described below and summarized in Table
\ref{tb.symm}.

At $T=0$ in the continuum one has the full rotational symmetry
$O(3)$ and the discrete symmetries of parity $P$ and charge conjugation
$C$.  States corresponding to the irreducible representations
(irreps) of $O(3)\times Z_2(P)\times Z_2(C)$ are labeled by $J^{PC}$.
On the lattice this group is broken to the discrete subgroup which
is the cubic group $O_h$. The $0^{PC}$ states (S/PS) become the
$A_1^{PC}$ of the cubic group and the $1^{PC}$ states (V/AV) become
the three dimensional irrep $F_1^{PC}$ (sometimes called the
$T_1^{PC}$) \cite{hamer}. The $2^{PC}$ breaks into a three dimensional
irrep called the $F_2^{PC}$ (or $T_2^{PC}$) and a two dimensional
irrep called the $E^{PC}$.

For $T\ne0$, the continuum theory has a $O(2)$ rotational symmetry
in the xy plane and a $Z_2({\cal T})$ symmetry (Euclidean time reversal)
in the Euclidean time direction. This constitutes the cylinder
group, $O(2)\times Z_2({\cal T})$, which is a subgroup of $O(3)\times
Z_2(P)$. On the lattice this is broken to the dihedral group $D_4^h
=D_4\times Z_2({\cal T})$ (where $D_4$ is the group of symmetries of a
square), which is also a subgroup of $O_h$, as is to be expected
\cite{dimred}.

The $J^{PC}$ irreps of $T=0$ break up into $M^{{\cal T}C}$ irreps
at finite temperature, where $M=J_z$ (since the cylinder group is
Abelian, its complex irreps are one-dimensional). The reduction
under the lattice symmetries is shown in Table \ref{tb.symm}. The
group theory is the most general. However, it does not restrict
special cases in which some of the independent representations
may become identical for dynamical reasons. We discuss these next.

Note that the $T=0$ vector breaks into two irreps for $T>0$.  This
group theory need not contradict our intuition that at sufficiently
low, but finite, temperature, the screening spectrum should be very
similar to the $T=0$ hadron spectrum, and therefore the thermal
$0^-$ and $1^+$ should be nearly degenerate. When the screening
mass, $\mu$, is sufficiently small (\ie, $\mu/T\gg1$), then the
effect of the boundaries must be exponentially small. The approximate
symmetry $O_h$ is broken to $D_4^h$ via terms that are exponentially
small, leading to near-degeneracy of the components of $T_1$ (the
splitting can only be observed with exponentially large statistics).
After all, the situation is not different from the $T=0$ theory
defined with various different boundary conditions--- they are
equivalent as long as $mL\gg 1$ where $L$ is the size of the box
in which the theory is defined.

However, this resolution raises another problem--- can the masses
of the two $0^-$ states, one obtained from the reduction of the PS
and the other from the V be equal? If the mass of the latter becomes
equal to that of the $1^+$ at low temperature, do we predict that
the PS and V states must be degenerate, contrary to previous
knowledge?  The answer must be negative, the resolution being that
the $0^-$ coming from the V cannot have overlap with the eigenvector
of the transfer matrix with the smallest eigenvalue. This argument
predicts that the screening masses in the V and PS channel (and
similarly in the S and AV channels) must be different, although the
naive transfer matrix argument gives a different result.

The correlator identities
\beq
   C_S(z) = -C_{PS}(z) \qquad C_V(z) = -C_{AV}(z)
\label{ids}\eeq
can be proven for overlap quarks by neglecting the effects of
instantons. This shows parity doubling. As can be seen from Table
\ref{tb.symm}, there is no group theoretical reason behind parity
doubling. If there were, then that would force chiral symmetry at
all $T$.

Gauge field configurations which give rise to chiral zero modes of
the overlap Dirac operator (called instantons) have interesting
consequences. Since these modes are localized, one recovers translation
invariance by averaging over many such configurations.  At high
temperatures it is observed that these configurations are rather
improbable. Hence, to take their effects properly into account one
must either use truly astronomical statistics or consider the zero
momentum correlators, $\chi_\Gamma$, \ie, correlators such as those
in eq.\ (\ref{corr}) summed over $1\le z\le N_z$. These have been
analyzed in \cite{earlier} and various identities were checked. One
should note that any configuration with a chiral zero mode breaks
the identity $C_S(z) = -C_{PS}(z)$. Configurations with zero modes
of both positive and negative chirality break the identity $C_V(z)
= C_{AV}(z)$. Thus, generic ensembles of gauge configurations
including instantons would realize only the irreps in Table
\ref{tb.symm}.  Parity doubling, represented in eq.\ (\ref{ids})
involves the disappearance of chiral zero modes. It has been noted
\cite{earlier} that these do not disappear at the QCD phase transition
but only gradually with increasing $T$. Hence parity doubling in
the high temperature phase is an approximate statement, and extremely
high statistics studies will be able to see the breaking of this
approximate symmetry.

\section{Free overlap quarks}

The overlap Dirac operator ($D$) can be defined \cite{neu2} in terms of the
Wilson-Dirac operator ($D_w$) by the relation
\beq
   D = 1 - D_w (D_w^\dag D_w)^{-1/2}.
\label{overlap}\eeq
When the gauge fields are translation invariant, $D_w$ can be diagonalized
in the Fourier basis. In free field theory, one can write
\beqa
\nonumber
   D_w &=& \a + i\gamma_\mu b_\mu,\qquad{\rm where}\\
   \a &=& 1+s-\sum_\mu\sin^2(p_\mu/2),\quad b_\mu=-\sin p_\mu,
\label{wilson}\eeqa
where $s$ is the (negative) mass parameter, and, at finite $T$, the
momenta are $p_0=2\pi(n+1/2)/N_t$ and $p_i=2\pi n/N_s$. Since $D^\dag
D = \a^2+b\cdot b$, and therefore diagonal,
the overlap Dirac operator is
\beqa
\nonumber
   D &=& \alpha + i \gamma_\mu \beta_\mu, \qquad{\rm where}\\
       \alpha &=& 1-\frac \a{\sqrt{\a^2+|b|^2}}, \quad
       \beta_\mu=\frac{b_\mu}{\sqrt{\a^2+|b|^2}},
\label{overlapdirac}\eeqa
where $|b|^2=b\cdot b$. Clearly the eigenvalues of $D$ lie on the
unit circle centered on the real axis at $(1,0)$.

Writing the eigenvalues of $D_w$ as $\lambda(D_w)=r\exp(i\theta)$,
with $r^2= \a^2+|b|^2$ and $\theta=\tan^{-1}(|b|/\a)$, it is easy to
see that the eigenvalues of $D$ are given by $\lambda(D)=\rho\exp(i\phi)$
where $\rho^2=2(1-\cos\theta)=2(1-\a/r)$ and $\phi=(\pi-\theta)/2$.
It is easy to check that in the limit $N_t\to\infty$ the minimum
eigenvalue is obtained by taking $p_0=\pi/N_t$ and $p_i=0$. This
gives
\beqa
   \rho &=& \frac\pi{(1+s)N_t}+{\cal O}\left(\frac\pi{N_t}\right)^3\\
   \phi &=& \frac\pi2+{\cal O}\left(\frac\pi{N_t}\right).
\label{mats}\eeqa
The factor of $1+s$ in the denominator is usually absorbed into a
redefinition of the fermion fields, thus giving the correct Matsubara
frequency in the continuum limit.

A massive overlap operator, $D(m)$, and the corresponding quark propagator,
$G(m)$, are defined by
\beqa
\nonumber
   D(m) &=& m + (1-m/2) D, \qquad{\rm and}\\
\nonumber
   G(m)&=&K^{-1}(m), \qquad{\rm where}\\
   K(m)&=&[1-D/2] D^{-1}(m),
\label{moverlap}\eeqa
and $m$ is the bare quark mass in lattice units \cite{neu2}.
The massive overlap propagator can, therefore, be written in the form
\beqa
\nonumber
   G(p;m) &=& f(p;m) + i\gamma_\mu h_\mu(p;m),\qquad{\rm where}\qquad\\
\nonumber
   f(p;m) &=& (1-\frac\rho2\cos\phi)(1+\frac m2)\cos\phi
     +\rho(1-\frac m2)\sin^2\phi,\\
\nonumber
   h_\mu(p;m) &=& \left[(1+\frac m2\cos\phi)\rho-(1-\frac m2\cos\phi)
     (1-\frac\rho2\cos\phi)\right]\\
    &&\qquad\qquad\times\sin\phi\beta_\mu.
\label{propm}\eeqa
As $m\to0$, we recover the usual massless propagator.

\begin{figure}[htb]\begin{center}
   \scalebox{0.45}{\includegraphics{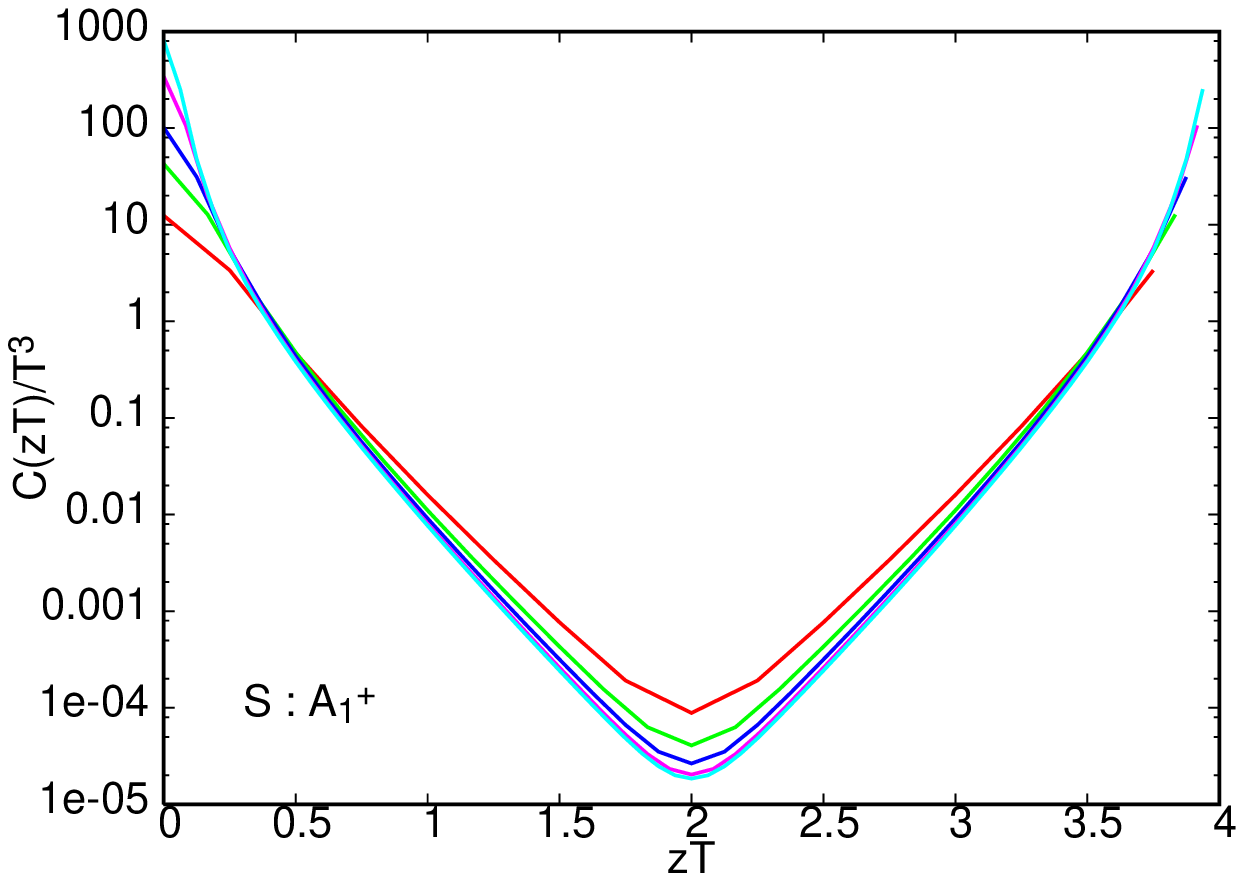}}
   \scalebox{0.45}{\includegraphics{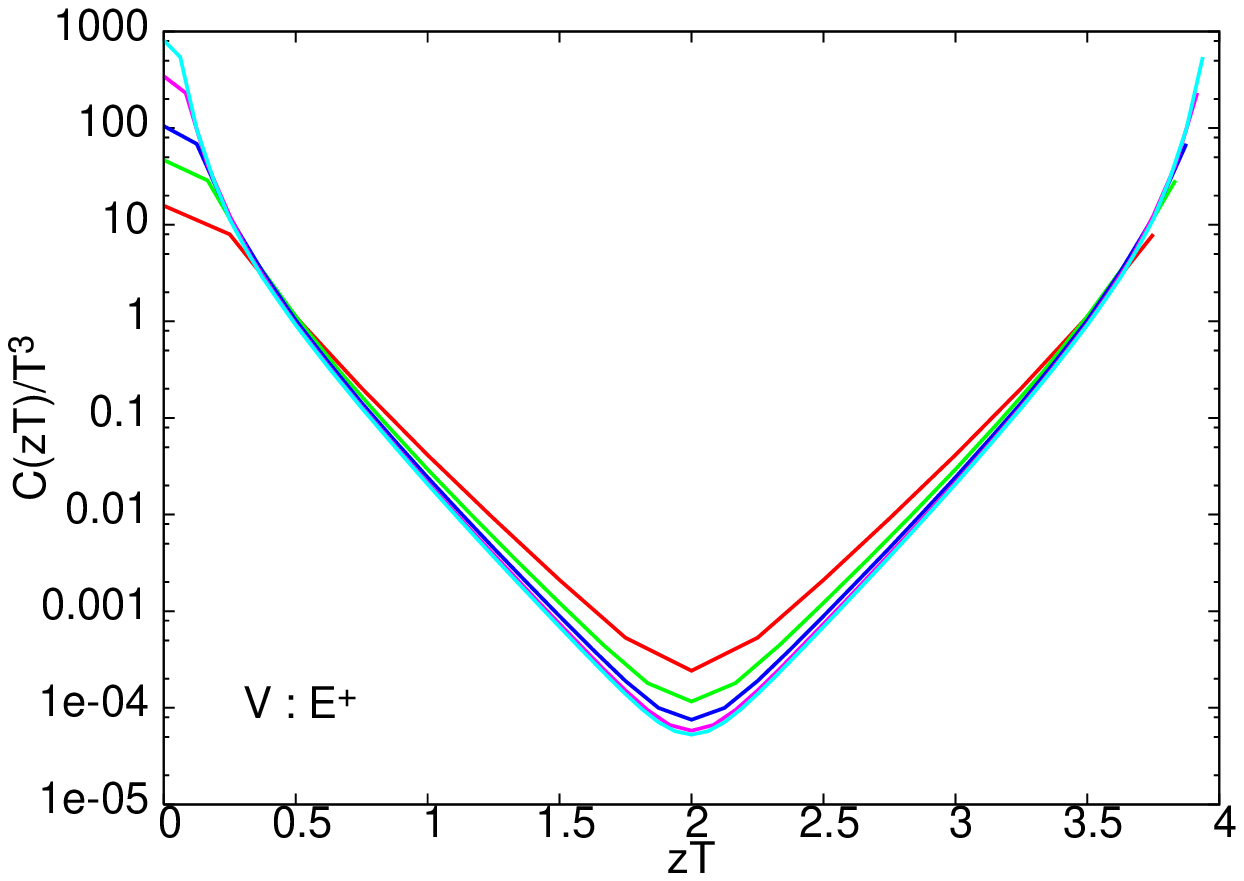}}
   \scalebox{0.45}{\includegraphics{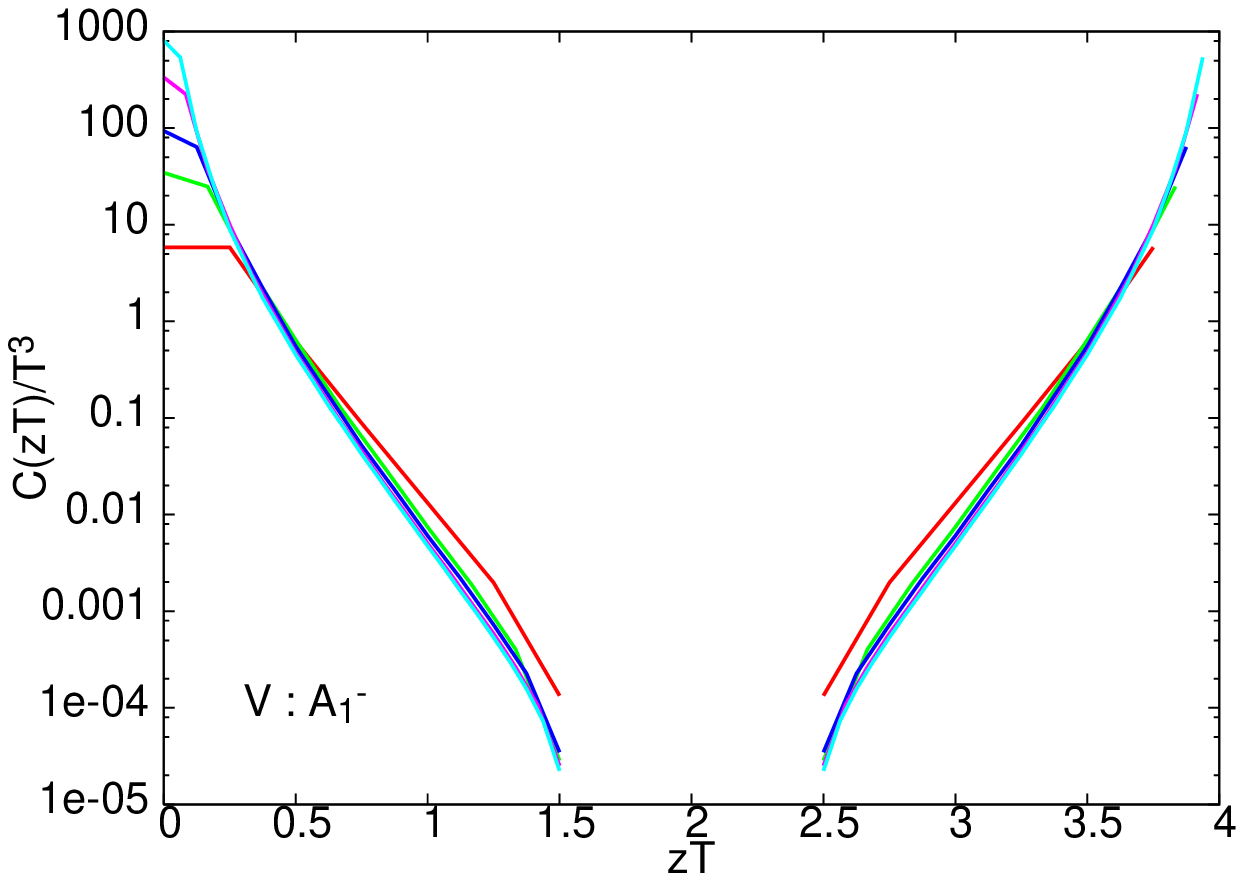}}
   \end{center}
   \caption{Correlators computed in a free field theory of overlap
    quarks on lattice sizes $4\times10^2\times16$, $6\times14^2\times24$
    $8\times20^2\times32$, $12\times30^2\times48$ and $16\times40^2\times64$.
    The $A_1^+$ correlator coming from
    S (first panel), the $E^+$ correlator coming from $V$ (second
    panel) and the $A_1^-$ correlator coming from $V$ (third panel)
    are shown. The value of $N_\tau$ increases from top to bottom
    at $zT=2$. The missing segments of the correlators in the last
    case correspond to the negative sign.}
\label{fg.fft}\end{figure}

The screening correlator at external momentum $q$ is then given by
\beq
   C_\Gamma(q) = \sum_p {\rm tr} G(p+q;m) \Gamma G^\dag(p;m)\Gamma.
\label{defcorr}\eeq
After performing the traces, one finds
\beq
   C_\Gamma(q) = 4\sum_p \left[ f(p+q;m) f(p;m) 
      + \Gamma_{\mu\nu} h_\mu(p+q) h_\nu(p)\right],
\label{mesons}\eeq
where the tensor $\Gamma_{\mu\nu}$ is $g_{\mu\nu}$ for S and
$2g_{\mu\lambda} g_{\nu\lambda}-g_{\lambda\lambda}g_{\mu\nu}$ for
the $\lambda$ polarization of V. The tensors $\Gamma_{\mu\nu}$ for
the PS and AV satisfy the correlator identities for overlap quarks.
Define $C_V$ to be the sum over three polarizations of the vector,
then with the Euclidean metric one finds $T_{\mu\nu}=-g_{\mu\nu}
-2g_{zz}$. As a result,
\beq
   C_V(q)+C_S(q) = -12 \sum_p g_z(p+q) g_z(p),
\eeq
which is generically non-zero.

In Figure \ref{fg.fft} we show some of the results of a numerical
computation of some of the meson-like correlators in free field theory
on lattices of various sizes. One sees a gradual convergence of the
results when plotted in terms of the scaling variables $zT=az/N_\tau$
and $C(zT)/T^3=a^3 C(az/N_\tau)/N_\tau^3$.  Several features of the
free field theory are interesting enough that we list them explicitly.
\begin{enumerate}
\item The correlator identities of eq.\ (\ref{ids}) are satisfied. This is
expected, since free field theory has no topological features which lead
to localized zero modes. As noted before, these identities are stronger
than the general feature expected from the group theoretical analysis,
since they allow us to equate the $A_1^+$ coming from PS to the $A+1^-$
coming from the S, \etc. 
\item The correlators in the $E^+$ and $A_1^-$ irreps coming from the V
have completely different behaviour. Consistent with this, the $E^-$ and
the $A_1^+$ correlators coming from the AV are quite different from each
other. However, the $E^\pm$ and the $A_1^\pm$ are identical.
\item The $A_1^-$ correlator from the V changes sign at $z\simeq2/T$.
The point at which the change of sign occurs is almost independent of
$N_z$ and $N_{x,y}$. This is clear evidence that a three dimensional
effective theory which describes these screening correlators cannot have
reflection positivity.
\item The $A_1^\pm$ coming from the V and AV are not degenerate with the
$A_1^\pm$ coming from the PS or the S, in agreement with the analysis of
the previous section.
\item The $A_1^\pm$ correlators from the PS and S are identical to the
$E^\pm$ correlators from the V and AV. There is no group theoretical
reason for this, and we shall examine later whether this holds in the
interacting theory.
\end{enumerate}

\section{Computational details}

Our numerical work was done in quenched QCD with overlap valence
quarks. We have computed all 8 components of the correlators
(one each for S and PS and three polarizations each for the
V and AV). Extensive finite size scaling studies were reported
earlier \cite{earlier} on lattices with $a=1/4T$. We have extended
this finite size scaling study here, but the main emphasis is
on studying the variation with lattice spacing and taking the
continuum limit. To this end we have used $a=1/6T$ and $1/8T$,
at the temperature $T=2T_c$, where $T_c$ is the critical temperature
for pure SU(3) gauge theory.

\begin{table}[hbtp]\begin{center}
  \begin{tabular}{|c|rrcc|c|}  \hline
  $\beta$ & Lattice Size & $N$ \\
  \hline

  $6.0625$ & $4\times10^2 \times 16$ &  19 \\
  $6.3384$ & $6\times14^2 \times 24$ &  20 \\
  $6.55$ & $8\times18^2 \times 32$ &  26 \\
  \hline
  $6.0625$ & $4\times12^3 $ &  23 \\
  $6.3384$ & $6\times14^3 $ &  16 \\
  \hline
  \end{tabular}
  \end{center}
  \caption[dummy]{Coupling $\beta$ for temperature $2T_c$, the lattice size, 
     and the number of configurations (separated by 1000 sweeps) used for 
     analysis ($N$). }
\label{tb.summary}\end{table}

The computation of propagators using the overlap Dirac operator of
eq.\ (\ref{overlapdirac}) needs a nested set of two matrix inversions
for its evaluation (each step in the numerical inversion of $D$
involves the inversion of $D_w^\dag D_w$).  This squaring of effort
makes a study of QCD with dynamical overlap quarks very expensive.
As before, we therefore chose to work with quenched overlap quarks.

We generated quenched QCD configurations at temperatures of $T/T_c$
= 2 on $4 \times 10^2 \times 16$, $6 \times 14^2 \times 24$ and $8
\times 18^2 \times 32$ lattices (see Table \ref{tb.summary}).  The
corresponding couplings are respectively $\beta= $ 6.0625, 6.3384
and 6.55.  Note that these are the known critical couplings on $N_t=
8$, 12 and 16 lattices.  For the latter, no infinite volume
extrapolation was available, unlike the first two. Our choice was
motivated by the 2-loop $\beta$-function, and consistency with the
the finite volume results. In each case, the configurations were
separated by 1000 sweeps of a Cabibbo-Marinari update.

For the matrix $M=D_w^\dag D_w$, and a given source vector $b$, we computed
$y=M^{-1/2} b$ by using the Zolotarev algorithm \cite{wupp}:  
\beq 
M^{-1/2} b =\sum_{l=1}^{N_O} \left(\frac{c_l}{M+d_l} b \right)~,~ \label{za}
\eeq 
where the coefficients $c_l$ and $d_l$ are computed with Jacobi
elliptic functions once one has the order of approximation ${N_O}$
and the ratio $\kappa=\mu_{\max}/\mu_{\min}$ where $\mu_{\max}$ and
$\mu_{\min}$ are the boundaries of the domain where we apply the
approximation.  In our implementation of the algorithm \cite{us2},
we first compute the lowest and highest eigenvalues of $M$ and then
choose $\mu_{\max}$ and $\mu_{\min}$ so that the domain of applicability
of the Zolotarev approximation is 10\% larger than the domain spanned
by the eigenvalues.  The order ${N_O}$ is defined by requiring  a
precision $\epsilon/2$ for the approximation of $1/\sqrt{z}$ in the
entire domain.  With these parameters, one calculates the approximation
in eq. (\ref{za}) by a multishift CG-inversion at the precision
$\epsilon/2$.

On each gauge configuration we computed $G$ on 12 point sources
(3 colors and 4 spins) for 8 quark masses from $m/T_c$=0.008 to 0.8
using a multi-mass inversion of $D^{\dagger}D$. The (negative) Wilson
mass term in $D_w$, \ie, $1+s$ in eq.\ (\ref{wilson}), which is an irrelevant regulator, was set to 1.8.
The tolerance ranged from $\epsilon=10^{-5}$ to $10^{-7}$ in the inner
CG and $10^{-3}$ to $10^{-5}$ in the outer CG, with most of the work
done using the larger values of $\epsilon$. We remark on the choice in
a later section. Typical $N_{\cal O}$ needed was about 7-8.

For most configurations we found that the spectrum of $D^\dag D$
starts well away from zero. However, very occasionally (one each
for $N_t=4$ and 8) we found a zero mode, \ie, an unpaired eigenvalue
$\lesssim10^{-2}$. As discussed in the literature, such zero modes have
to be subtracted off to obtain averages of physical quantities such as
the correlation functions. We showed elsewhere that our determination of
the eigenvectors and eigenvalues is precise enough that we can subtract
out the zero mode contribution in the chiral condensate. The resultant
subtracted condensate (or correlators) is well within the statistical
distribution of the measurement obtained in the sample without zero
modes, at all the couplings and lattice sizes studied.  We therefore
chose not to include the zero mode contributions below.  Furthermore,
we show below results for our lowest quark mass $m/T_c$ =0.008, as the
main features are essentially quark mass independent.

\section{Results}

\subsection{Symmetry realization}

\begin{figure}[htb]\begin{center}
   \scalebox{0.65}{\includegraphics{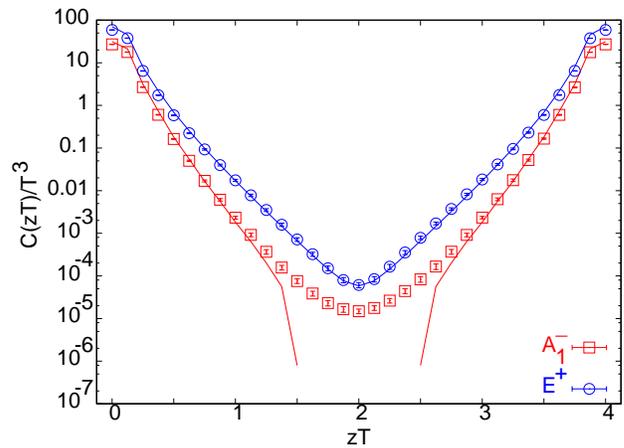}}
   \end{center}
   \caption{The $A_1^-$ and the $E^+$ correlators coming from the
    V correlator on an $8\times18^2
    \times 32$ lattice at $T=2T_c$. Also shown are the results of a
    free field theory (FFT) computation.}
\label{fg.symm}\end{figure}

\begin{figure}[htb]\begin{center}
   \scalebox{0.65}{\includegraphics{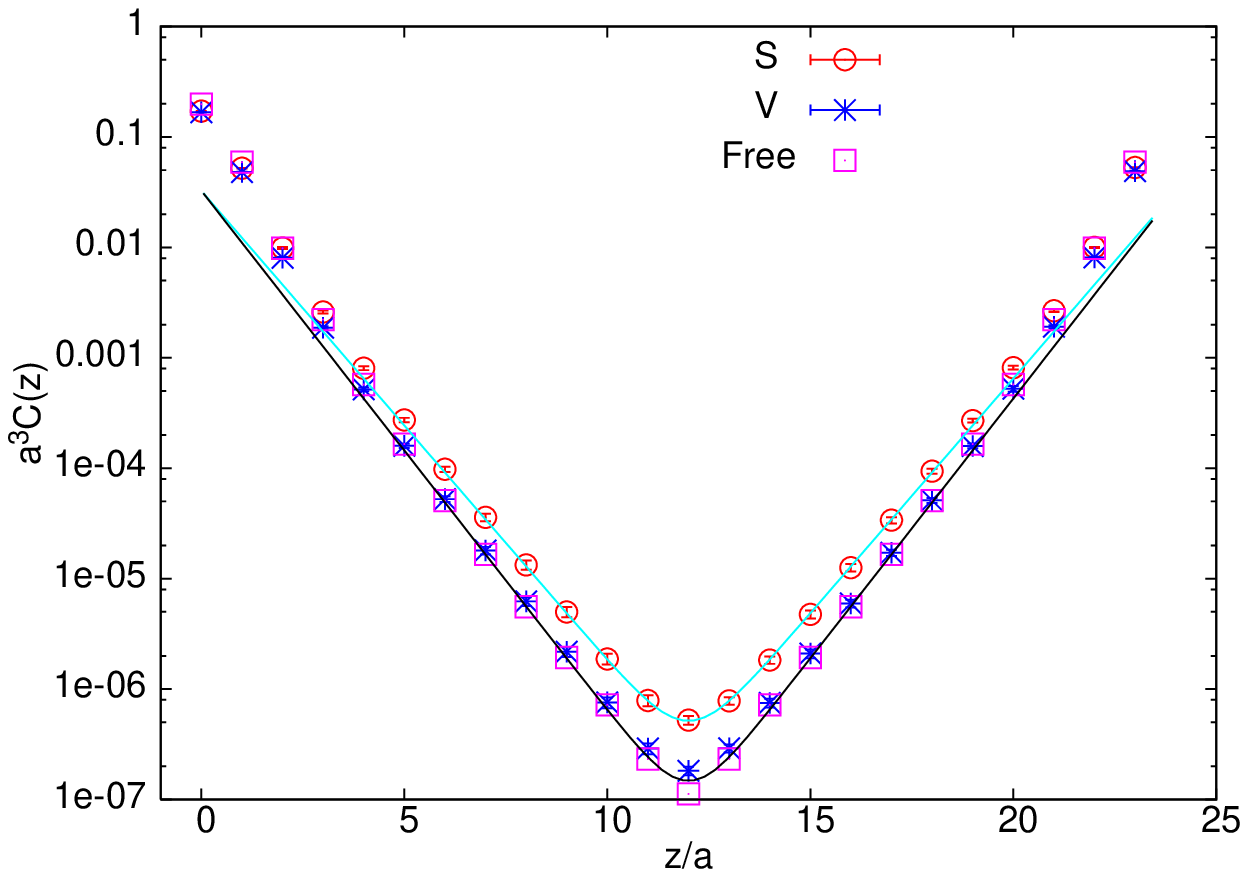}}
   \scalebox{0.65}{\includegraphics{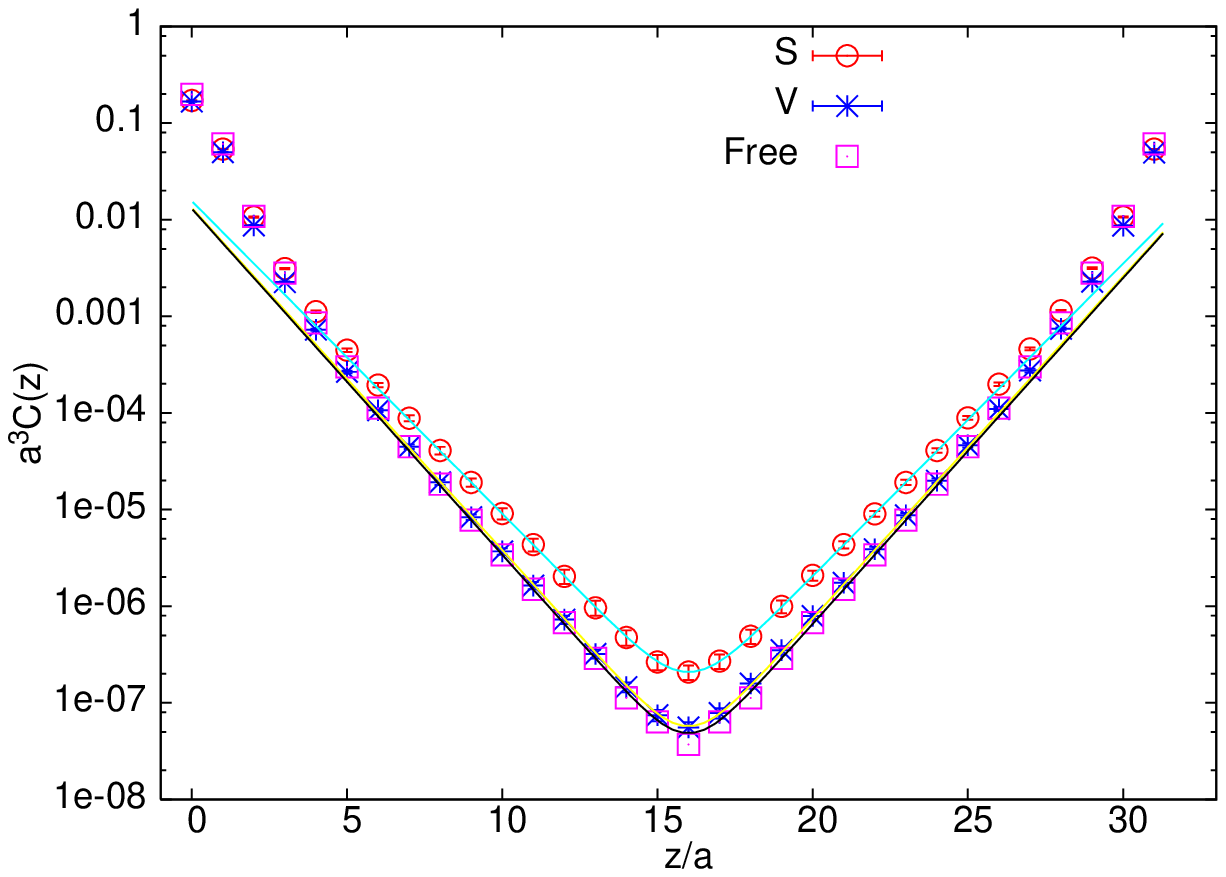}}
   \end{center}
   \caption{Screening correlators in the S and V channels compared
    with the corresponding ideal gas results on a $6\times14^2 \times
    24$ lattice (first panel) and an $8\times18^2 \times 32$ lattice
    (second panel) at $T=2T_c$. The lines are single $cosh$ fits.}
\label{fg.corr}\end{figure}

\begin{figure}[htb]\begin{center}
   \scalebox{0.65}{\includegraphics{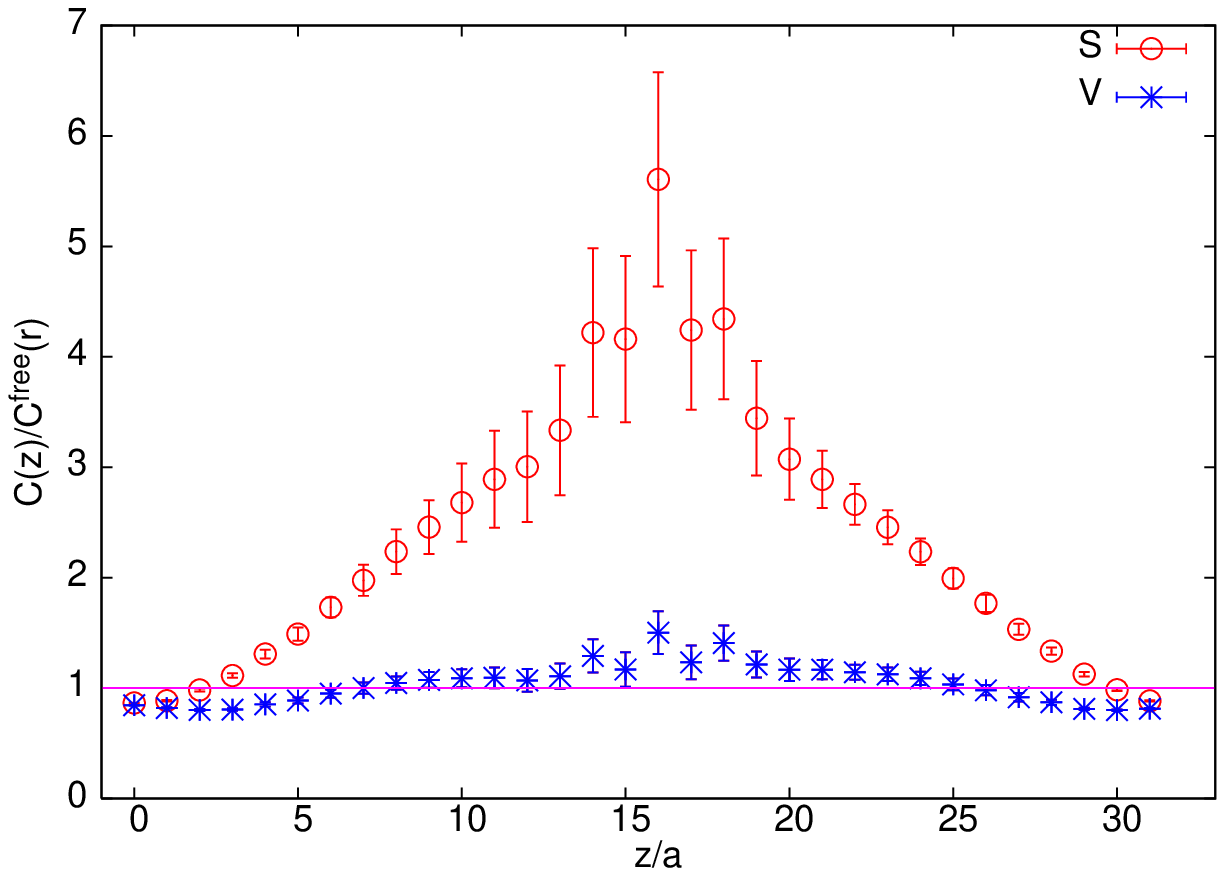}}
   \end{center}
   \caption{The ratio of the measured and ideal gas screening correlators
    in the S and V channels.}
\label{fg.ratid}\end{figure}

\begin{figure}[htb]
   \begin{center}
   \scalebox{0.65}{\includegraphics{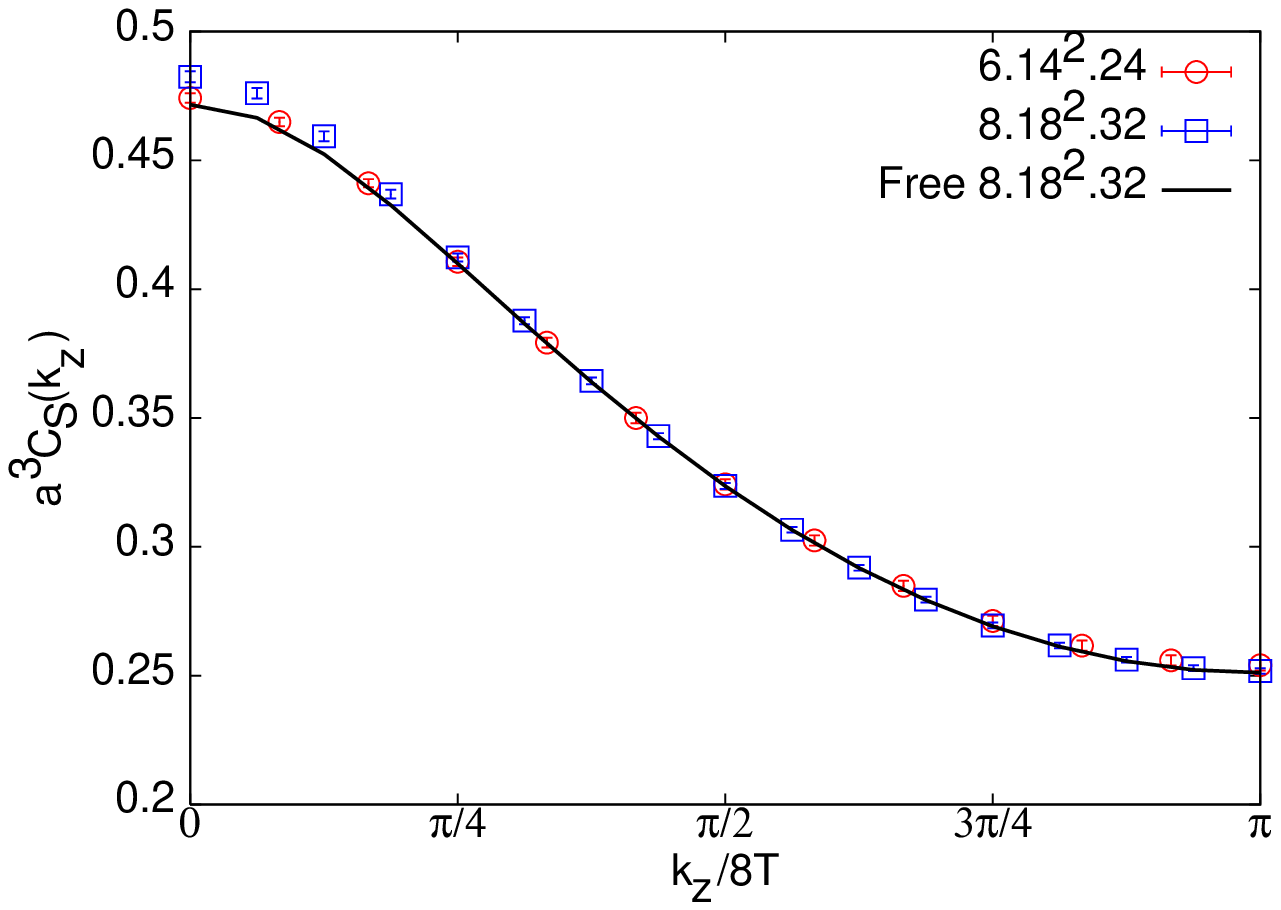}}
   \scalebox{0.65}{\includegraphics{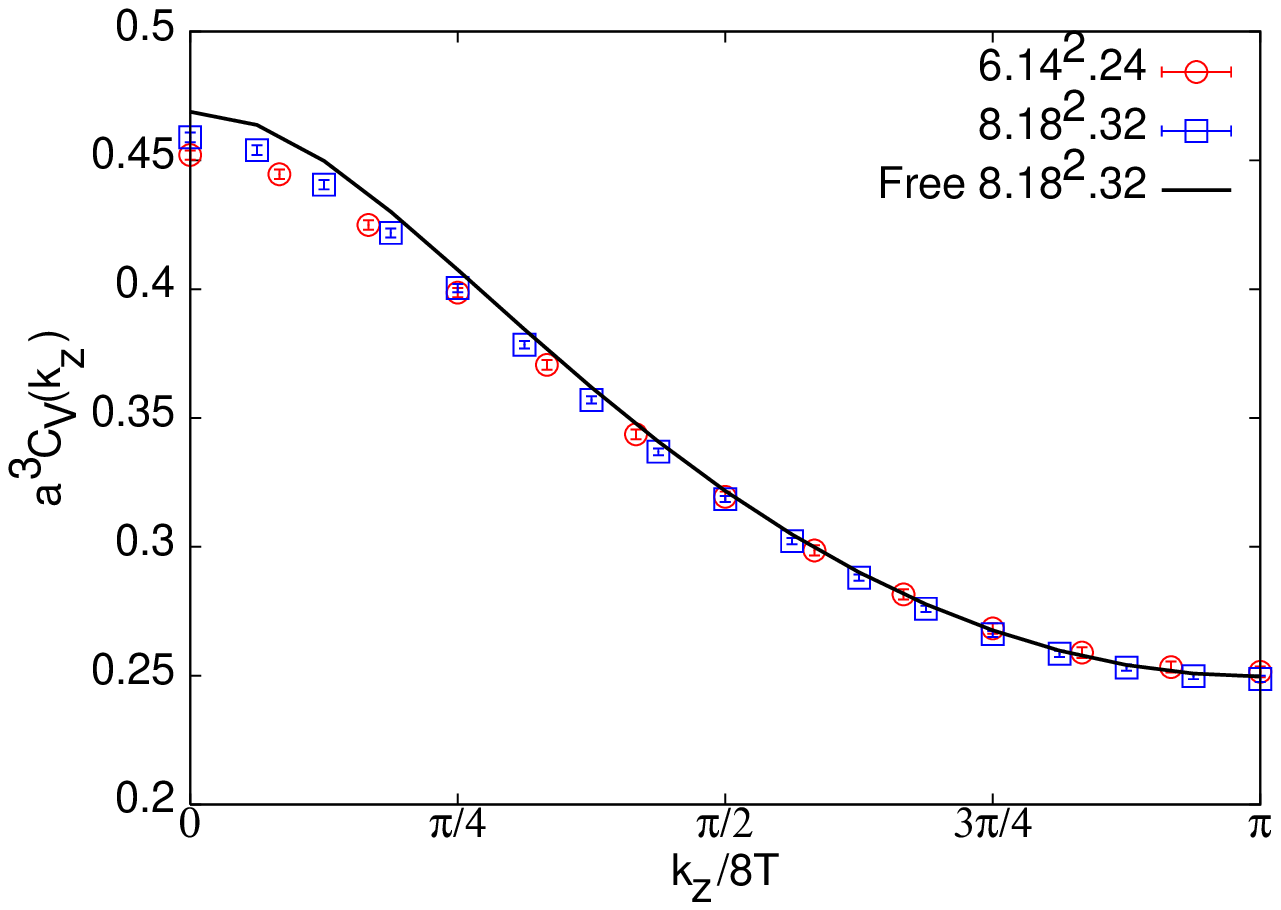}}
   \end{center}
   \caption{Momentum space correlators at two different lattice spacings
      compared with the ideal gas results, normalized to fit the
      high-momentum data, for the S (first panel) and V
      (second panel) channels.}
\label{fg.mom}\end{figure}

We have checked that the sum of the S and PS correlators vanishes within
errors in all our measurements, in agreement with the correlator
identities for overlap quarks.  For the V and AV correlators we
checked that the difference of the $x$ and $y$ polarization correlators
vanish within errors, thus verifying that they lie in the same irrep
of the symmetry group, as indicated by Table \ref{tb.symm}.

In Figure \ref{fg.symm} we show the screening correlator obtained from
various polarizations of the V. The $t$-direction polarization
gives the $A_1^-$ and the $x$ and $y$ directions give the $E^+$. In the
figure we show the correlator from the sum of the $x$ and $y$ polarizations.
We also show the corresponding correlator obtained from a free field
theory (FFT) computation. Note the close agreement between the $E^+$
correlator in the the quenched theory and the FFT results, over six
orders of magnitude change in the correlator. We will examine this
in more detail later.

The disagreement between the FFT and the full theory for the $A_1^-$
correlator shown in Figure \ref{fg.symm} requires further analysis.
When the tolerance in the fermion inversion was decreased, the $E^+$
correlator changed only marginally, whereas the $A_1^-$ correlator
changed sign at $zT>3/2$ on some gauge configurations. However, further
work is required to check whether these correlators also turn negative
as the tolerance is decreased further.

In the past the sum of the three polarizations of the $T=0$ vector
have been analyzed together. Strictly speaking, this is incorrect,
since it mixes the $E^\pm$ and $A_1^\mp$ irreps. However, the screening
correlator in the latter irrep is almost an order of magnitude smaller
than that in the former. As a result, the screening masses extracted
by the older method is a fairly accurate measure of the screening mass
in the $E^\pm$ channel. In fact, we find that the screening mass from
the $E^+$ correlator at $N_t=8$ is $a\mu_{E^+}=0.798(4)$ whereas that
from the mixed representation is $a\mu_V=0.816(4)$. The two screening
masses are compatible with each other at the $3\sigma$ level.

With this caveat in mind,
in the remainder of this paper we shall report results obtained with
the older method of analysis, where the V correlator sums over all the
polarizations. It has the advantage that the results are directly comparable
with earlier results, and is not far removed from the quantitatively
correct results which would be obtained by following the theoretically
correct method.

\subsection{Correlations as functions of spatial separation}

In Figure \ref{fg.corr} we compare the measured correlation functions
in the S and V channels. It is noteworthy that the V
correlator agrees rather well with the ideal gas correlation functions
from $z=0$ all the way to $N_z/2$, when the correlator itself falls
by 6--7 orders of magnitude. To make this spectacular point more
forcefully, we have plotted in Figure \ref{fg.ratid} the ratio of
the measured correlator with the ideal gas. It is clear that the
V ($E^+$) correlator differs from the ideal gas by no more than
10-20\% everywhere except (possibly) near the center of the lattice.

The situation is somewhat different for the S ($A_1^+$) correlator.
This agrees well with the ideal gas up to a distance $z\simeq1/T$.
At distances larger than this, the S channel correlator falls
significantly slower than the ideal gas result (Figure \ref{fg.corr}).
This is shown again in Figure \ref{fg.ratid}, where it becomes clear
that the ratio of the correlators changes by a factor of about 5
when the correlators themselves have changed by about 6 orders of
magnitude.

\subsection{Correlations as functions of momentum}

\begin{figure}[htb]\begin{center}
   \scalebox{0.65}{\includegraphics{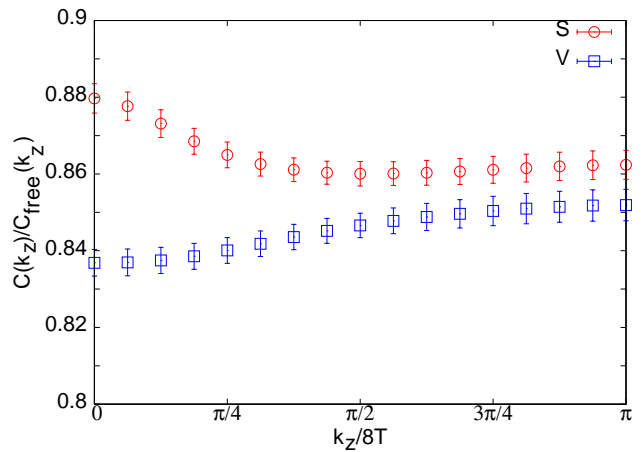}}
   \end{center}
   \caption{The ratio of the measured and ideal gas screening correlators
    in the S and V channels, on the $8.18^2.32$ lattice.}
\label{fg.momrat}\end{figure}

In some applications the momentum space correlator is a more direct
input, giving emphasis to the small $z$ region. Our lattice data can be exhibited in momentum space by a
Fourier transformation of the separation $z$. The periodic boundary
conditions give rise to a Brillouin zone structure as usual, and
only momenta in the range $0\le k_z\le \pi TN_t$ are independent.

We show the momentum space correlators in Figure \ref{fg.mom}. It
turns out that the ideal gas result needs to be multiplied by a
constant (independent of $k_z$) in order to describe the data well.
However, after this normalization, which we choose to fix the
correlator at $k_z=\pi TN_t$, the shapes of the correlators are in
rough agreement with the ideal gas results, except at small $k_z$.
From Figure \ref{fg.mom} another interesting feature which is visible
is that the correlators at two different lattice spacings, $a=1/6T$
and $1/8T$, after the scaling by $a^3$, show no significant dependence
on lattice spacing at $k_z=\pi/4a$, $\pi/2a$, $3\pi/4a$ and $\pi/a$.
However, there are small differences visible at low momenta, $k_z<\pi
T$, for both the S and V correlators. With decreasing $a$,
these residual $a$-dependent effects go in the direction of slightly
improving the agreement with free field theory for V, and of
making matters slightly worse for S.

The small disagreements with free field theory can be better exhibited
by displaying the ratio of the correlator and its ideal gas
counterpart- $C_\Gamma(k_z)/C_\Gamma^{free}(k_z)$. These ratios are
shown in the V and S channels for the $8.18^2.32$ lattice in
Figure \ref{fg.momrat}.  The ratio shows clear structure. Near the
center of the Brillouin zone, $k_z\approx4\pi T$, the ratios go to
a momentum independent constant.  However, this constant is different
from unity, and also different in the two channels. For $k_z<2\pi
T$ the ratio is clearly momentum dependent.  In the S channel,
the ratio decreases with $k_z$ whereas in the V channel it
increases.  To the best of our knowledge there is no prediction or
understanding of such structures in weak coupling theory.

\subsection{Screening masses}

\begin{figure}[htb]\begin{center}
   \scalebox{0.65}{\includegraphics{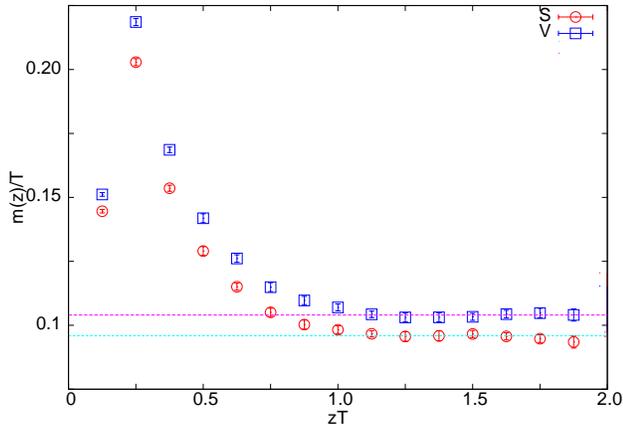}}
   \end{center}
   \caption{Effective masses extracted on the $8\times18^2\times32$
    lattice, compared with the results of the fits to a single cosh
    indicated by the horizontal lines.}
\label{fg.localm}\end{figure}

\begin{figure}[htb]\begin{center}
   \scalebox{0.65}{\includegraphics{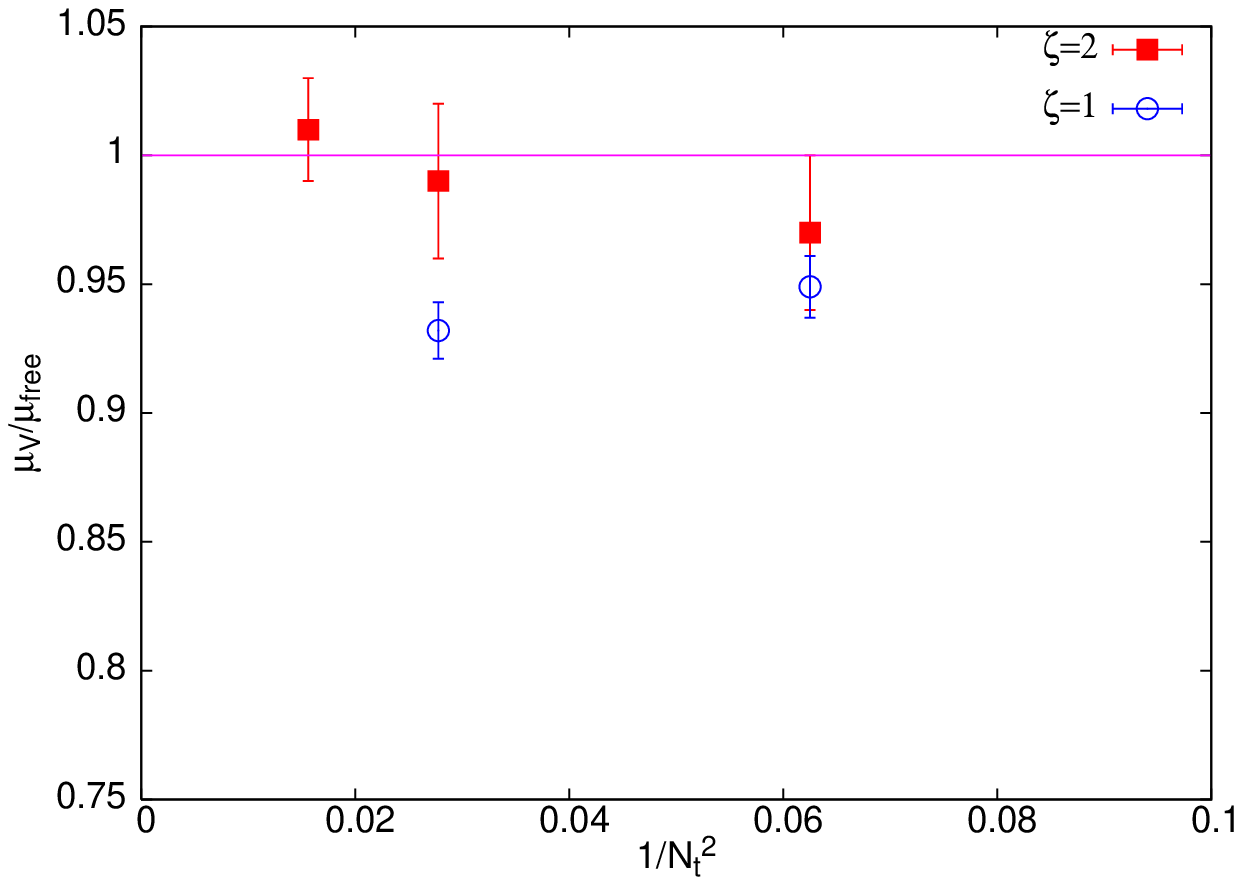}}
   \scalebox{0.65}{\includegraphics{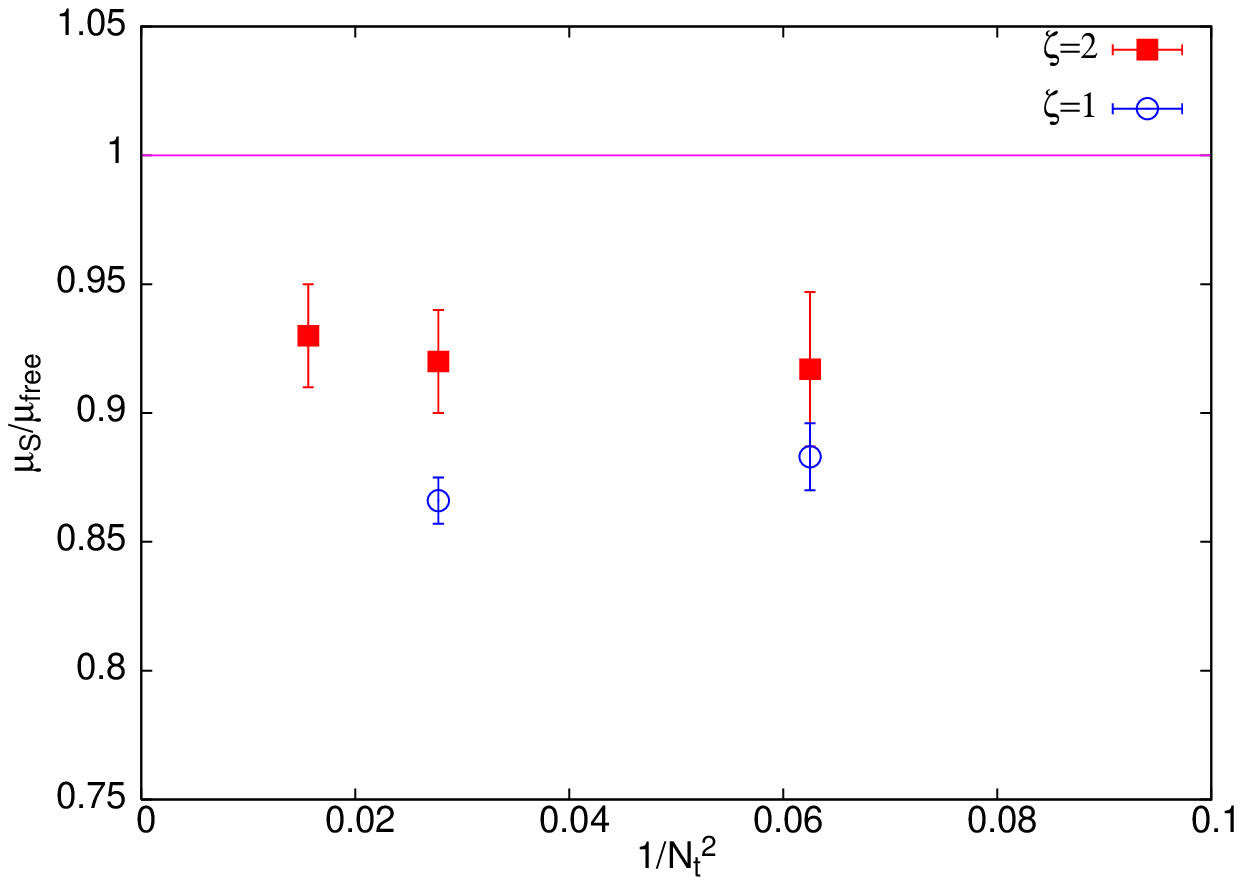}}
   \end{center}
   \caption{The ratio of the screening mass in quenched QCD to that
    in a theory of free overlap quarks on the same lattice for the
    S and V channels at $T=2T_c$. Results for two aspect
    ratios, $\zeta=1$ and 2, are shown.}
\label{fg.scrm}\end{figure}

\begin{figure}[htb]\begin{center}
   \scalebox{0.65}{\includegraphics{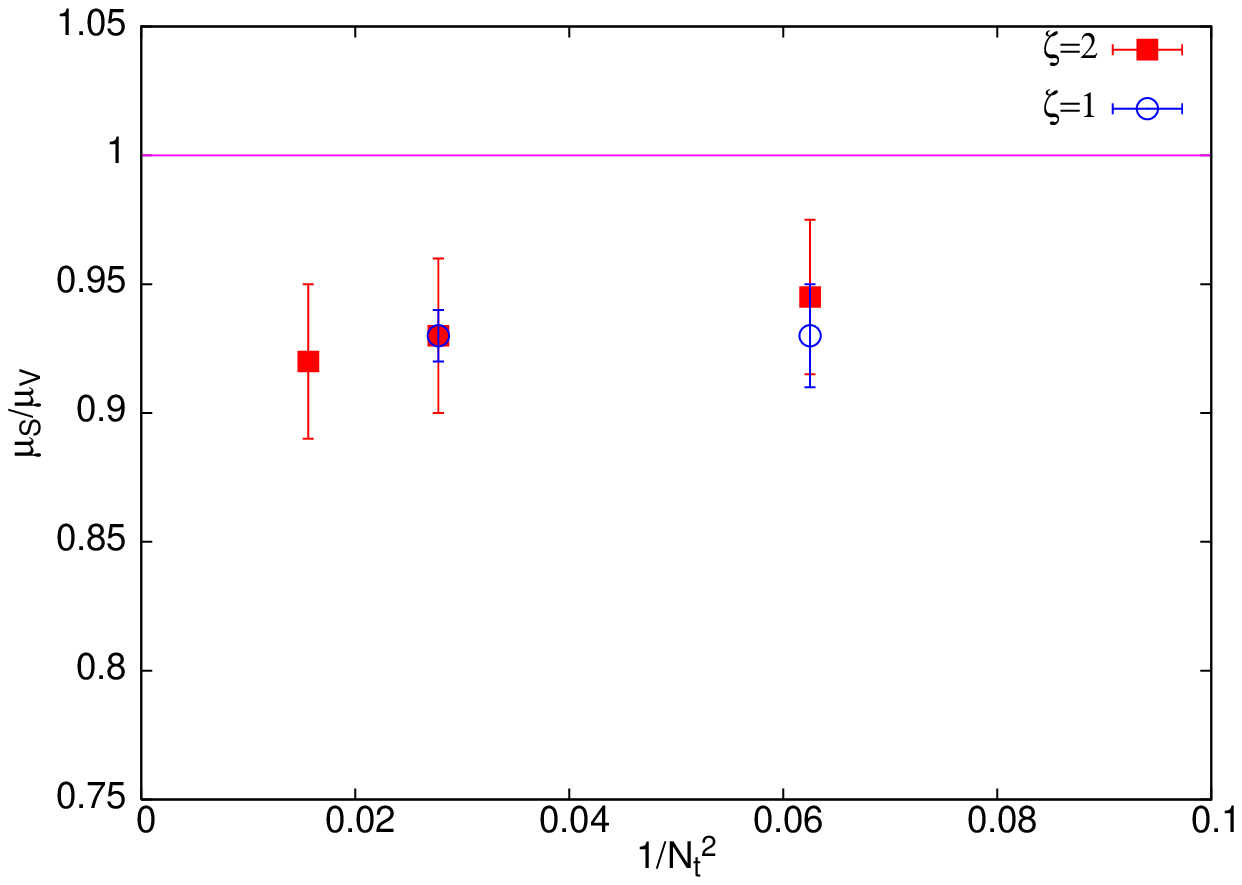}}
   \end{center}
   \caption{The ratio of the screening mass in the S and V
    channels in quenched QCD are shown as a function of $1/N_t^2\propto a^2$
    for $T=2T_c$. Results for two aspect ratios, $\zeta=1$
    and 2, are shown.}
\label{fg.scrcomp}\end{figure}

One can extract screening masses from the correlators in two different
ways. One is by fitting a single cosh to the long distance part of
the screening correlator. The other method is to use the ratio
$C(z)/C(z+1)$ and the assumption that at distance between $z$ and
$z+1$ is described by a single cosh to extract a distance dependence
effective mass, $m(z)$. If there is a clear plateau in the effective
mass as a function of $z$, which agrees with the fitted value, then
one could reliably talk of a screening mass and its value.

In Figure \ref{fg.localm} we exhibit the result of such a test. It
is clear that there is a stable plateau over almost half the available
distances, and the local mass in this region ($z>1/T$) agrees well
with the results of a fit over the same range. This is quite different
from earlier studies of staggered and Wilson quarks, where a stable
plateau in the local masses was not observed. Note also the need for
aspect ratios $\zeta=N_z/N_s>1$ in order to stabilize the local masses
at their asymptotic value.

Notice the non-monotonicity of the local masses at the smallest distance.
If the correlation function could be expressed in the spectral form---
\beq
   C(z) = \sum_i A_i\left[\exp(-m_i z) + \exp(-m_i\{N_z-z\})\right],
\eeq
with all the $A_i\ge0$, \ie, the correlation function came from a
transfer matrix satisfying reflection positivity, then the local masses
would be a non-increasing function of distance for $z<N_z/2$. Clearly,
therefore, the data shows that this condition is violated. Such behaviour
has been observed with both staggered and Wilson quarks earlier. Note
also the lack of convexity of the screening correlators in Figure
\ref{fg.corr}, which is also a consequence of the violation of reflection
positivity.

In Figure \ref{fg.scrm} we plot the ratio of screening masses
extracted in quenched QCD to that in a free field theory of overlap
quarks on the same lattice, $\mu/\mu_{\mathrm free}$, as a function of
$1/N_t^2\propto a^2$ at fixed $T=2T_c$. In the V channel the
ratio is consistent with unity. In the S channel, however, the
ratio differs from unity at the 95\% confidence level but is almost
independent of the lattice spacing.

We have shown results for two values of the aspect ratio, $\zeta=1$
and 2. At the smallest lattice spacing we have only $\zeta=2$. It would
be useful to extend these computations to larger $\zeta$. If
reflection positivity of the screening correlators were to hold,
then one would expect the screening masses not to increase with
$\zeta$.  However, in Figure \ref{fg.scrm} we see that this exactly
what happens, again indicating that reflection positivity is violated.

In view of our results, it is interesting to ask whether it is
possible that at some larger value of $N_t$ there is a cross over
to a regime where $\mu_S$ is closer to $\mu_{\mathrm free}$. If
such were the case, then one should be able to identify the physics
of such a cross over. There is only one such piece of physics, and
that is the scale of non-locality of the overlap Dirac operator.
This has been studied in \cite{nonlocal}, where it was shown that
the scale of non-locality varies smoothly towards zero as the lattice
cutoff decreases. This allows us to rule out a cross over from the
observed behaviour of $\mu_S$ to trivial physics.

A direct comparison of the screening masses in the S and V
channels is shown in Figure \ref{fg.scrcomp}. From this it is clear that
these two masses differ from each other at the 95\% confidence limit
at the smallest lattice spacing that we use, and show no tendency
to move towards equality. Since the ratio $\mu_S/\mu_V$ is within
5\% of unity, an explanation within a weak coupling expansion seems
possible. However, there is no such explanation available at this
time.

\section{Summary}

We have performed a group theoretical analysis of the screening
transfer matrix at finite temperature. Quark bilinear operators,
in the $C=+1$ sector, which are classified as S, PS, V and AV at
zero temperature can be decomposed at finite temperature as shown
in Table \ref{tb.symm}. The thermal state $A_1^-$ is obtained in
the decomposition of both the PS and V. Parity doubling is not a
consequence of the group theory, but arises dynamically. Some
identities which can be proven for the topologically trivial sector
of overlap quarks, eq.\ (\ref{ids}), show parity doubling.  We
presented an argument that if the PS and V are not to be degenerate
at low temperatures, then they cannot be degenerate at any temperature.
The theory of free overlap quarks is consistent with the converse,
since it gives a degeneracy between the PS and V at all temperature.
As a result, this theory cannot be an accurate representation of all
aspects of QCD at finite temperature.

We have demonstrated---
\begin{enumerate}
\item The correlation function in the $E^+$ sector (V) is very
   close to that in free field theory. In the $A_1^+$ sector (S)
   there is a significant difference at large distances, \ie, for
   $z>1/T$. However, the magnitude of such differences are small
   compared to the observed fall of the correlation function by
   6 orders of magnitude over the distances involved.
\item Both the S and V correlators are very close to the
   prediction of free field theory at distance less than $T$. 
\item The screening mass of the V is consistent
   with the free field theory of overlap quarks, whereas that of the
   S is lighter by about 5\% at the 95\% confidence limit. This
   behaviour persists into the continuum limit.
\item The S and V correlators in momentum space agree
   with free field theory of overlap quarks, after an overall normalization,
   only near the center of the Brillouin zone. The part with $k_z<\pi T$
   differs from free field theory for both the S and the V.
\item There is evidence that the screening transfer matrix does not have
   reflection positivity. This implies
   that screening in finite temperature QCD cannot be equivalent to the
   zero temperature physics of a temperature dependent Hamiltonian. This
   statement is not controversial, since screening phenomena at finite
   temperature crucially involve a mixed state density matrix, whereas
   correlations at zero temperature involve pure quantum states.
\end{enumerate}

\noindent
None of these features depend upon topological non-triviality of the gauge
field ensemble, since the one that has been used in these computations have
vanishing topology.

This work was funded by the Indo-French Centre for the Promotion of 
Advanced Research under its project number 3104-3. These computations were
performed on the Cray X1 of the Indian Lattice Gauge Theory Initiative
(ILGTI) in TIFR, Mumbai.


\begin{thebibliography}{99}
\bibitem{linkage}
  R.\ V.\ Gavai and S.\ Gupta, {\sl Phys.\ Rev.\/} D 73 (2006) 014004.
\bibitem{precise}
  S.\ Gupta, {\sl Phys.\ Rev.\/} D 64 (2001) 034507.
\bibitem{nadkarni}
  S.\ Nadkarni, {\sl Phys.\ Rev.\/} D 33 (1986) 3904;
    {\sl ibid\/} D 34 (1986) 3904.
\bibitem{yaffe}
  P.\ Arnold and L.\ G.\ Yaffe, {\sl Phys.\ Rev.\/} D 52 (1995) 7208.
\bibitem{hels}
  K.\ Kajantie \etal, {\sl Phys.\ Rev.\ Lett.\/} 79 (1997) 3130.
\bibitem{saumen}
  S.\ Datta and S.\ Gupta, {\sl Phys.\ Rev.\/} D 67 (2003) 054503.
\bibitem{jpsi}
  S.\ Datta \etal, {\sl Phys.\ Rev\/} D 69 (2004) 094507.
\bibitem{detar}
  C.\ DeTar and J.\ Kogut, {\sl Phys.\ Rev.\ Lett.\/} 59 (1987) 3784.
\bibitem{mtc}
  K.\ D.\ Born {\sl et al.\/}, {\sl Phys.\ Rev.\ Lett.\/} 67 (1991) 302.
\bibitem{quenched}
  A.\ Gocksch {\sl et al.\/}, {\sl Phys.\ Lett.\/} B 205 (1988) 334;
  S.\ Gupta, {\sl Phys.\ Lett.\/} B 288 (1992) 171.
\bibitem{two}
  S.\ Gottlieb {\sl et al.\/}, {\sl Phys.\ Rev.\ Lett.\/} 59 (1987) 1881,
  {\sl Phys.\ Rev.\/} D 47 (1993) 3619, {\sl ibid.\/} D 55 (1997) 6852;
  J.\ B.\ Kogut {\sl et al.\/}, {\sl ibid.\/} D 58 (1998) 054504.
\bibitem{four}
  G.\ Boyd {\sl et al.\/}, {\sl Z.\ Phys.\/} C 64 (1994) 331;
  R.\ V.\ Gavai and S.\ Gupta, {\sl Phys.\ Rev.\ Lett.\/} 85 (2000) 2068.
\bibitem{wilson}
  T.\ Hashimoto {\sl et al.\/}, {\sl Nucl.\ Phys.\/} B 400 (1993) 267;
  Ph.\ de Forcrand {\sl et al.\/}, {\sl Phys.\ Rev.\/} D 63 (2001) 054501;
  E.\ Laermann and P.\ Schmidt, hep-lat/0103037.
\bibitem{symm}
  S.\ Gupta, {\sl Phys.\ Rev.\/} D 60 (1999) 094505.
\bibitem{pos}
  R.\ V.\ Gavai, S.\ Gupta and R.\ Lacaze {\sl PoS\/} Lat2006 (2006) 135.
\bibitem{tifr}
  R.\ V.\ Gavai and S.\ Gupta, {\sl Phys.\ Rev.\ Lett.\/} 83 (1999) 3784.
\bibitem{conti}
  R.\ V.\ Gavai and S.\ Gupta, {\sl Phys.\ Rev.\/} D 67 (2003) 034501.
\bibitem{edwin}
  S.\ Wissel \etal, {\sl PoS LAT2005\/}, (2006) 164.
\bibitem{neu}
  H.\ Neuberger and R.\ Narayanan, {\sl Phys.\ Rev.\ Lett.\/} 71 (1993) 3251.
\bibitem{luescher}
  M.\ L\"uscher, {\sl Phys.\ Lett.\/} B 428 (1998) 342.
\bibitem{dimred}
  S.\ Datta and S.\ Gupta, {\sl Nucl.\ Phys.\/} B 534 (1998) 392.
\bibitem{hamer}
  M.\ Hamermesh, {\sl Group theory and its application to physical problems\/},
   Addison-Wesley, Reading MA, USA (1962).
\bibitem{laine}
  M.\ Laine and M.\ Veps\"al\"ainen, {\jhep} 02 (2004) 004.
\bibitem{neu2}
 H.\ Neuberger, {\sl Phys.\ Lett.\/} B 417 (1998) 141.
\bibitem{earlier}
  R.\ V.\ Gavai, S.\ Gupta and R.\ Lacaze,
    {\sl Phys.\ Rev.\/} D 65 (2002) 094504.
\bibitem{nonlocal}
  P.\ Hernandez, K.\ Jansen and M.\ L\"uscher,
    {\sl Nucl.\ Phys.\/} B 552 (1999) 363.
\bibitem{wupp}
  J.\ van den Eshof et al., {\sl Comput.\ Phys.\ Comm.\/} 146 (2002) 203.
\bibitem{us2}
  R.\ V.\ Gavai, S.\ Gupta and R.\ Lacaze, {\sl Comput.\ Phys.\ Comm.\/}
    154 (2003) 143.
\end{thebibliography}
\end{document}